\documentclass{article}
\usepackage{amsmath,amsthm,amssymb,tikz,url}
\usetikzlibrary{arrows}

\title{Fault-Tolerant, but Paradoxical Path-Finding in Physical and Conceptual Systems}
\author{Bryan Knowles and Mustafa Atici\\
Western Kentucky University\\
Bowling Green, KY, USA}
\begin{document}
	\maketitle
	\begin{abstract}
		We report our initial investigations into reliability and path-finding based models and propose future areas of interest. Inspired by broken sidewalks during on-campus construction projects, we develop two models for navigating this ``unreliable network." These are based on a concept of ``accumulating risk" backward from the destination, and both operate on directed acyclic graphs with a probability of failure associated with each edge. The first serves to introduce and has faults addressed by the second, more conservative model. Next, we show a paradox when these models are used to construct polynomials on conceptual networks, such as design processes and software development life cycles. When the risk of a network increases uniformly, the most reliable path changes from wider and longer to shorter and narrower. If we let professional inexperience--such as with entry level cooks and software developers--represent probability of edge failure, does this change in path imply that the novice should follow instructions with fewer ``back-up" plans, yet those with alternative routes should be followed by the expert?
	\end{abstract}
	
	\section{Introduction}
	A few years ago, Western Kentucky University began a series of construction projects, recently completing Gary A. Randsdell Hall, the Augenstein Alumni Center, and nearing completion of Downing Student Union, to replace the Downing University Center at the heart of the main campus. Part of this reconstruction process is the periodic and temporary removal of lengths of sidewalks throughout campus, allowing workers to access lines--water, electrical, etc.--underneath the concrete.
	
	To minimize the disruption, these removals are scheduled for the less-busy Summer and Winter semesters, and it was during one such Summer that I paid my rent by working on campus. Each morning, I would bike from my apartment near the Southwest edge of campus to the Northeast edge where I worked as a Web Application Developer for the Kentucky Mesonet.
	
	Quickly, though, I became frustrated of backtracking around these broken sidewalks, eventually posing the question, ``What's the \emph{most reliable path} on campus?"
	
	\section{Review}
	This problem has obvious and well researched implications for computer and information networks, but it also has implications for design processes, introduced in section \ref{s:imp}. We note here the research done on related graph theory and fault-tolerance problems and their differences with our problem.
	
	Thomas Wolle considers an undirected graph with risk probabilities associated with each edge, computing after multiple failures ``for a large number of different properties Y whether Y holds" \cite{wolle02}. Akiyuki Yano and Tadashi Wadayama consider undirected Erdos-Renyi random graphs with risk probabilities associated with each edge, setting an upper and lower bound to the probability that the network will become disconnected \cite{yano12}. Allen Chang and Eyal Amir consider directed acyclic graphs where a risk is associated with each edge, but these risks are functions of a hidden random variable \cite{chang12}.
	
	Panagiotis Papadimitratos et al and Weidong Cui et al seek multiple paths through a network such that the correlation of failures between the paths is minimized \cite{papadimitratos02,cui02}. Pitu Mirchandani and Anthony Chen, in separate works, consider directed graphs with uncertain distances between nodes, seeking the ``expected shortest travel time" \cite{mirchandani76} or an optimal path through the network \cite{chen05}. Kayi Lee et al consider multilayer networks, examining the relationships between failures across levels \cite{lee11}.
	
	Another problem, given as an assignment in an algorithms course \cite{kosaraju13}, asks for the most reliable path when rerouting around failed edges is not allowed. Its solution is simple: if $S$ is the set of edge success rates, select a path that minimizes $\Sigma_i log S_i$.
	
	But although our problem is like each of these in part, it is unlike all of them in whole. We seek, on arbitrary, simple, directed, and acyclic graphs, a dynamic route that minimizes the likelihood of a lone pathfinder having to backtrack when each edge has a random chance of failure; a path whose back-up paths have back-up paths, \emph{ad infinitum}; a fault-tolerant path-finding \emph{strategy}.
	
	\section{Definitions}
	Our solution is based on the concept of ``accumulating risk" backward from the destination vertex. Once the source vertex has been reached, the pathfinder greedily chooses at each step the edge with the least accumulated risk. When a failed edge has been encountered, the pathfinder updates its knowledge of the network, re-accumulates, and continues.
	
	To model this behavior, let the pathfinder's current knowledge of the network be the graph $G=(V,J)$ with a set of risk probabilities $R$ associated with the set of edges $J$. Let $R_i$ denote the risk of edge $i$ and $\hat{R}_i$ its accumulated risk.
	
	Our principle thesis then is determining the proper definition of $\hat{R}$ such that a greedy pathfinder behaves optimally.
	
	\subsection{Eagle-Eye Model}
	We refer to our first definition for $\hat{R}$ as the ``eagle-eye" model. This model, whose risks and accumulated risks are denoted respectively with $E$ and $\hat{E}$, is based on intuition and expressed in equation \ref{eq:E}, where $N(i)$ is the set of outgoing neighbors of $i$.
	
	\begin{equation}
		\hat{E}_i = E_i \vee \Pi_{j \in N(i)} \hat{E}_j
		\label{eq:E}
	\end{equation}
	
\begin{figure}
		\centering
		\begin{tikzpicture}
			\tikzset{vertex/.style={shape=circle,draw=black,scale=0.9}}
			\tikzset{edge/.style={->,> = latex}}
			\tikzset{label/.style={fill=white,scale=0.9}}
			\node[vertex] (n1) at (0,1) {1};
			\node[vertex] (n2) at (2,2) {2};
			\node[vertex] (n3) at (3,0) {3};
			\node[vertex] (n4) at (4,3) {4};
			\node[vertex] (n5) at (4,1) {5};
			\node[vertex] (n6) at (6,3) {6};
			\node[vertex] (n7) at (6,1) {7};
			\node[vertex] (n8) at (6,0) {8};
			\node[vertex] (n9) at (8,1) {9};
			\foreach \from/\to/\w in {n1/n2/.81,n1/n3/.88,n2/n4/.78,n2/n5/.78,n3/n8/.75,n4/n6/.75,n4/n7/,n5/n6/.75,n5/n7/.75,n6/n9/.50,n7/n9/.50,n8/n9/.50}
				\draw[edge] (\from) to node[label]{\w} (\to);
		\end{tikzpicture}
		\caption{Approximate $\hat{E}$ values for a simple graph. Each edge is assumed to have $E=0.50$. \label{fg:EG}}
	\end{figure}
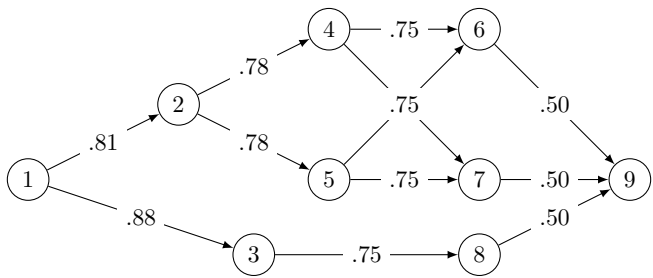

	Imagine the chance that a path through edge $i$ does \emph{not} fail. Both $i$ must succeed and a successful path must exist through at least one of its neighbors. Equation \ref{eq:E} is simply the logical inverse of this statement, expressing risk instead of success. Figure \ref{fg:EG} gives an example.
	
	However, this model can ``see too far" down the sidewalk. Imagine that edge $i$ leads into an infinite number of neighbors, each with a high, but less than certain, chance of failure $\alpha$. Almost certainly there exists at least one of those infinite edges that leads to a successful path to the destination vertex, so all that needs to be worried about then is the risk that $i$ itself will fail--at least that's how this model treats it. Equation \ref{eq:En} shows this.
	
	\begin{equation}
		\lim_{n \to \infty} \hat{E}_i = E_i \vee \alpha^n = E_i \vee 0 = E_i
		\label{eq:En}
	\end{equation}
	
	But this is not how our pathfinder operates! If backtracking were allowed or if our pathfinder had ``eagle-eyes" allowing it to see down the entire length of a sidewalk, around buildings, and through obstacles, then this model might satisfice. However, with backtracking forbidden, once our pathfinder selects an edge, it has committed itself to that edge. It seems then that the proper definition of $\hat{R}$ must have knowledge of how the pathfinder selects edges.
	
	\subsection{Bat-Eye Model}
	The ``bat-eye" model takes into account the pathfinder's limited range of sight when navigating the network and \emph{the order} at which it will attempt to traverse edges. That is, the neighbors of $i$ have been sorted, in some manner of speaking, from least to greatest accumulated risk.
	
	This model, whose risks and accumulated risks are denoted respectively with $B$ and $\hat{B}$, is defined by equation \ref{eq:B}. $C(i;j)$ is the probability that neighbor $j$ will be chosen from $i$ and $A(j)$ is the accumulated risk \emph{remaining after} edge $j$, defined respectively in equations \ref{eq:C} and \ref{eq:A}. Because $C(i;j)$ requires knowledge of the order edges will be attempted by the pathfinder, we let $N(i;j)$ denote the neighbors of $i$ attempted before $j$.
	
	\begin{equation}
		\hat{B}_i = B_i \vee [\Pi_{j \in N(i)} B_j + \Sigma_{j \in N(i)} C(i;j) A(j)]
		\label{eq:B}
	\end{equation}
	
	\begin{equation}
		C(i;j) = (1 - B_j) \Pi_{k \in N(i;j)} B_k
		\label{eq:C}
	\end{equation}
	
	\begin{equation}
		A(j) = \frac{\hat{B}_j - B_j}{1 - B_j}
		\label{eq:A}
	\end{equation}
	
	Again imagine the chance that a path through edge $i$ does not fail. First, $i$ must succeed, then at least one neighbor must succeed, and finally whatever lies ahead of the chosen neighbor must succeed. Equations \ref{eq:B}-\ref{eq:A} together are the logical inverse of this statement. Note the use of a weighted sum to take advantage of the disjunction between selecting a next step.
	
	Unlike its predecessor, this model is not fooled by infinite neighbors. As in the last model, the product will converge to zero, but the summation here will be a weighted average of the risk after the infinite neighbors. Equation \ref{eq:Bn} shows this, where it's assumed without loss of generality that $A(j)=\beta$.
	
	\begin{equation}
		\lim_{n \to \infty} \hat{B}_i = B_i \vee [\alpha^n + \Sigma C(i;j) \beta] = B_i \vee [\alpha^n + \beta (1-\alpha^n)] = B_i \vee \beta
		\label{eq:Bn}
	\end{equation}
	
	\section{Validation}
	If the combined risk of all paths going through each edge is known, why shouldn't the pathfinder choose that which minimizes this risk? Equation \ref{eq:B} determines this combined risk and can be broken into three terms: the first, $B_i$, represents the probability that $i$ itself will fail; the second, $\Pi B_j$, the probability that all of the neighbors of $i$ will fail; and the third, $\Sigma C(i;j) A(j)$, a summation of disjoint probabilities, each the chance that a given neighbor will be chosen and the risk remaining after that edge.
	
	\begin{equation}
		\hat{R}_i =
		\begin{cases}
			R_i & \text{Terminal} \\
			R_i \vee [\Pi R_j + \Sigma C(i;j) A(j)] & \text{Otherwise} \\
		\end{cases}
		\label{eq:R}
	\end{equation}
	
	Equation \ref{eq:R} extends equation \ref{eq:B} with a base case--if an edge ends at the destination, i.e. is terminal, no accumulation need take place; note that as a consequence, $A(i)=0$ for all terminal edges $i$. Proof of correctness is straightforward. $\hat{R}=R$ is trivially true for terminal edges. For edges leading only into terminal edges, the third term will be ``zeroed out" and the remaining two terms, identical here to the eagle-eye model, are again trivially true. All other edges can be said to lead into a mix of distances from the destination, where the three terms together cover the full range of possibilities recursively. Finally, because ``bat-eye" defines ``risk" as it would be experienced by a greedy pathfinder on a directed acyclic graph, these are by definition the values with which it behaves optimally.
	
	\section{Implications} \label{s:imp}
	We now turn our attention to the likely scenario where $R$ is unknown. Unless there is specific information on the distribution of lengths and risks in the network, we can only assume all edges are of equal length and have equal risk $0 \leq \alpha \leq 1$. If left as a variable, this has an interesting result when substituted for $R_i$ in either model: a polynomial will be associated with each edge \cite{lee11}. For convenience, let $P(\alpha; i)$ denote the polynomial for edge $i$ when $R=\{\alpha, ...\}$.
	
	\begin{figure}
		\begin{tikzpicture}
			\tikzset{vertex/.style={shape=circle,draw=black,scale=0.9}}
			\tikzset{edge/.style={->,> = latex}}
			\node[vertex] (n1) at (0,1) {1};
			\node[vertex] (n2) at (1,2) {2};
			\node[vertex] (n3) at (1,0) {3};
			\node[vertex] (n4) at (2,3) {4};
			\node[vertex] (n5) at (2,1) {5};
			\node[vertex] (n6) at (3,3) {6};
			\node[vertex] (n7) at (3,1) {7};
			\node[vertex] (n8) at (3,0) {8};
			\node[vertex] (n9) at (4,1) {9};
			\foreach \from/\to in {n1/n2,n1/n3,n2/n4,n2/n5,n3/n8,n4/n6,n4/n7,n5/n6,n5/n7,n6/n9,n7/n9,n8/n9}
				\draw[edge] (\from) to (\to);
		\end{tikzpicture}
		\includegraphics[width=0.4\textwidth]{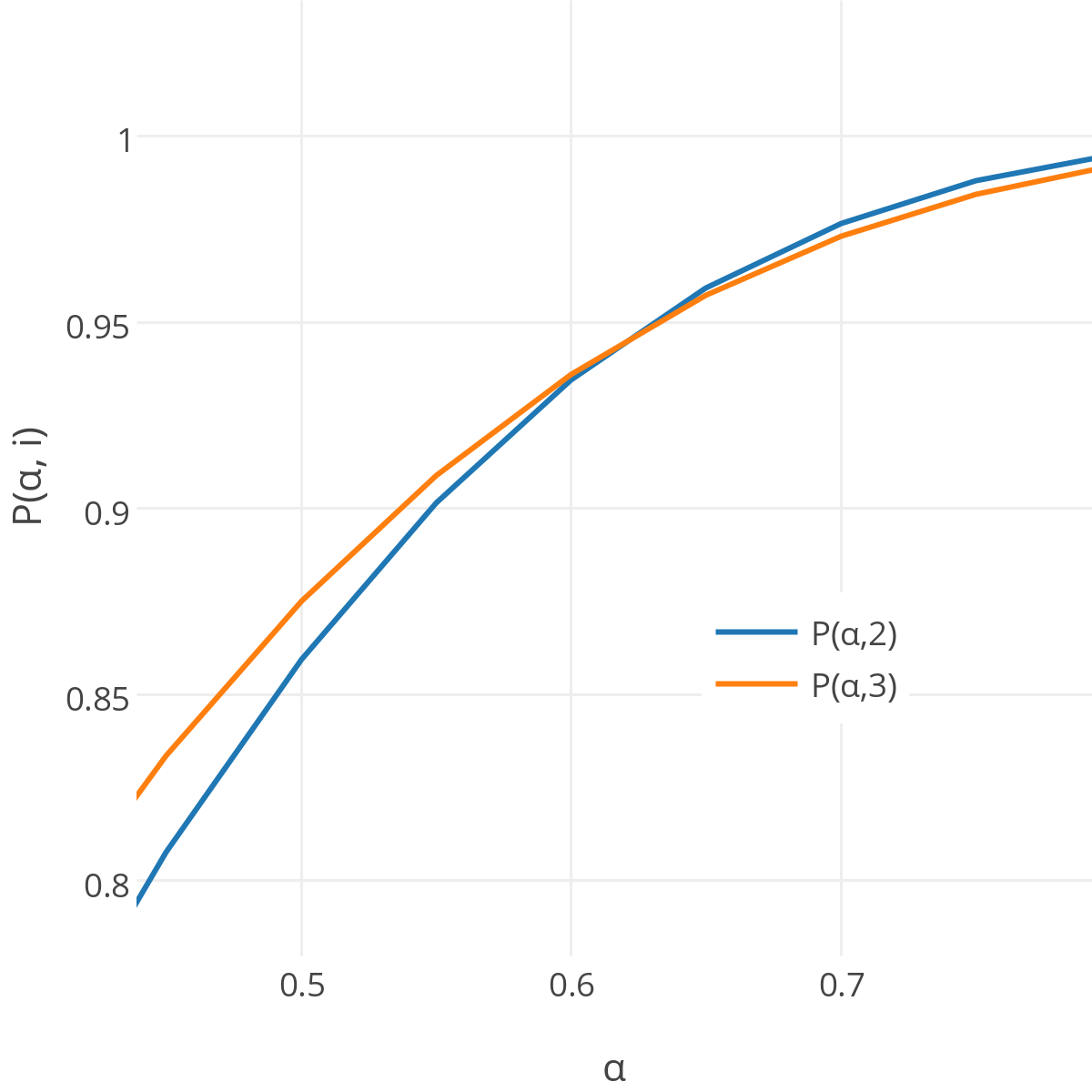}
		\begin{equation*}
			P(\alpha, 1\rightarrow2) = -\alpha^6 + 2\alpha^5 + \alpha^4 -4\alpha^3 + \alpha^2 +2\alpha,
			P(\alpha, 1\rightarrow3) = \alpha^3 - 3\alpha^2 +3\alpha
		\end{equation*}
		\caption{A small graph exhibiting the ``risk paradox," where node one is the source and node nine is the destination. Polynomials for the first two choices are given. \label{fg:G}}
	\end{figure}
	
	The consequence of these polynomials in development applications is a ``risk paradox" where no universally safest path exists. That is, as the value of $\alpha$ is tuned from zero to one, the path that would be followed by a greedy pathfinder changes. In the example of figure \ref{fg:G}, when $\alpha<0.6180$ it is more reliable to take the longer, but denser path; above that value, however, it is best to ``just run for it" on the shortest path. Furthermore, as the following examples show, the optimal path one determines for certain actors is contrary to intuition.
	
	\subsection{Baking Cakes}
	An accessible example of this paradox could be following instructions to make a cake. Instead of risks being associated with each edge, risks are associated with each chef's experience. Therefore an expert chef would have a low $\alpha$ whereas a novice's would be high. And instead of a straightforward recipe, there are alternative steps throughout the process that ultimately lead to the same cake.
	
	It may seem that the less experienced chef may wish to take the conceptual route with more ``back-up" plans. But in a scenario structured like figure \ref{fg:G} with a success rate of one-third or less, those back-up plans are not enough to offset the increased number of steps.
	
	\subsection{Software Engineering}
	A hypothetical software firm has developed a design process structured as a graph that exhibits the ``risk paradox." This firm has contracts with clients who expect products delivered reliably, and the employees of this firm have a wide range of experience levels.
	
	A developer intern has been assigned a particularly difficult solo project--perhaps as a test of skill. What route through the development process should he or she set out to follow? Should this novice concern him- or herself with the ``most reliable shortest path" \cite{kosaraju13}? Should the start-up ``just run for it" until they've gathered the necessary experience to work together as a team? How should the project manager consider skill levels when planning the order of tasks?
	
	Because this paradox--and this problem--have little applicability in processes that are straightforward or have minimal branching, perhaps we will benefit by including in our consideration explorative phases, such as the initial conversations with the client. Solidifying requirements following those meetings will again be straightforward, albeit possibly iterative if the client has uncertain or unclear needs, so we look further to the next explorative phase in the typical software development life cycle, planning.

	During this phase, a touch of artfulness is required to develop a set of system components that will produce the desired set of effects. The firm may have its conventions on how its engineers should proceed, but these are simply skeletons or springboards or defaults; decisions still must be made by humans about test conditions and object interactions! The ability to delineate options in these situations, we posit, will differ noticeably when trying to do so quickly versus when trying to do so consistently.
	
	Is it better to develop software in as few strokes of genius as possible or to ensure that strokes of genius will continue to be had until the project is complete and the client is happy?
	
	\section{Conclusion and Future Work}
	It cannot be overlooked that there are complexities in the design process. Herbert Simon proposed a curriculum for a ``design science," stepping through a syllogism starting with the philosophical definition of ``artifact" and ending with a model of man that framed all human activity as navigation through conceptual space \cite{simon96}. Under this view, where design proceeds by examining possible next steps and carrying them out in order and seeking some satisfactory end-goal, our path-finding problem is particularly powerful, providing the basis for a framework of ``reliable design."
	
	Our research into this area is in its infancy. In communication networks, it has been thoroughly explored \cite{chen12}. We concern ourselves then with physical systems that closely resemble our ``broken sidewalk" inspiration--if not literally sidewalks, where their condition in cities has degraded to the extent that ``even if the sidewalks miraculously stopped breaking, at the current pace it would take 69 years to repair all the existing damage" \cite{shoup09}. Second, we will consider conceptual systems in which workflows can be characterized as directed and acyclic, perhaps as a ``branching waterfall process." Finally, we plan to explore network analysis metrics based on reliability models, such as an alternative measure of ``betweeness centrality."
	
	\bibliographystyle{plain}
	\bibliography{references.bib}

\begin{thebibliography}{10}

\bibitem{chang12}
Allen Chang and Eyal Amir.
\newblock Reachability under uncertainty.
\newblock {\em arXiv preprint arXiv:1206.5253}, 2012.

\bibitem{chen05}
Anthony Chen and Zhaowang Ji.
\newblock Path finding under uncertainty.
\newblock {\em Journal of advanced transportation}, 39(1):19--37, 2005.

\bibitem{chen12}
Biyu Chen.
\newblock Reliable shortest path problems in networks under uncertainty:
  models, algorithms and applications.
\newblock 2012.

\bibitem{cui02}
Weidong Cui, Ion Stoica, and Randy~H Katz.
\newblock Backup path allocation based on a correlated link failure probability
  model in overlay networks.
\newblock In {\em Network Protocols, 2002. Proceedings. 10th IEEE International
  Conference on}, pages 236--245. IEEE, 2002.

\bibitem{kosaraju13}
S.~Rao Kosaraju.
\newblock Solution to assignment 8.
\newblock http://bit.ly/1pa0m2X, 2013.

\bibitem{lee11}
Kayi Lee, Hyang-Won Lee, and Eytan Modiano.
\newblock Reliability in layered networks with random link failures.
\newblock {\em IEEE/ACM Transactions on Networking (TON)}, 19(6):1835--1848,
  2011.

\bibitem{mirchandani76}
Pitu~B Mirchandani.
\newblock Shortest distance and reliability of probabilistic networks.
\newblock {\em Computers \& Operations Research}, 3(4):347--355, 1976.

\bibitem{papadimitratos02}
Panagiotis Papadimitratos, Zygmunt~J Haas, and Emin~G{\"u}n Sirer.
\newblock Path set selection in mobile ad hoc networks.
\newblock In {\em Proceedings of the 3rd ACM international symposium on Mobile
  ad hoc networking \& computing}, pages 1--11. ACM, 2002.

\bibitem{shoup09}
Donald Shoup.
\newblock Putting cities back on their feet.
\newblock {\em Journal of Urban Planning and Development}, 136(3):225--233,
  2009.

\bibitem{simon96}
Herbert~Alexander Simon.
\newblock {\em The sciences of the artificial}.
\newblock MIT press, 1996.

\bibitem{wolle02}
Thomas Wolle.
\newblock A framework for network reliability problems on graphs of bounded
  treewidth.
\newblock In {\em Algorithms and Computation}, pages 137--149. Springer, 2002.

\bibitem{yano12}
Akiyuki Yano and Tadashi Wadayama.
\newblock Probabilistic analysis of the network reliability problem on a random
  graph ensemble.
\newblock In {\em Information Theory and its Applications (ISITA), 2012
  International Symposium on}, pages 327--331. IEEE, 2012.

\end{thebibliography}
\end{document}